\documentclass[a4]{sig-alternate-10pt}
\usepackage[T1]{fontenc}
\usepackage{lmodern}
\usepackage[utf8]{inputenc}
\usepackage[unicode]{hyperref}
\usepackage{xspace}
\usepackage{subfig}
\usepackage{caption}
\usepackage{color}
\usepackage{graphicx}
\usepackage{booktabs}
\usepackage{multirow}
\usepackage{tabularx}
\usepackage[group-digits=true, group-separator={,}]{siunitx}
\usepackage[pointedenum]{paralist}
\usepackage{url}
\usepackage{listings}

\usepackage{algorithmic}
\usepackage{algorithm}

\newcommand{\var}[1]{\textit{#1}}
\algsetup{linenodelimiter=}

\newcommand{\specialcell}[2][c]{%
  \begin{tabular}[#1]{@{}c@{}}#2\end{tabular}}

\newif\ifstatus
\statustrue

\hypersetup{pdfstartview=FitV,
  colorlinks=true, linkcolor=black, citecolor=black, urlcolor=blue,
  pdftitle={New kids on the blocks: an analysis of modern blockchains}, pdfauthor={TBA}, pdfkeywords={TBA}
}

\newcounter{nalg}[section] 
\renewcommand{\thenalg}{\thesection .\arabic{nalg}} 
\DeclareCaptionLabelFormat{algocaption}{Algorithm \thenalg} 

\newcommand{\latinlocution}[1]{\textit{#1}}

\newcommand{\eg}{\latinlocution{e.g.,}\xspace}
\newcommand{\ie}{\latinlocution{i.e.,}\xspace}
\newcommand{\etal}{\latinlocution{et al.}\xspace}

\begin{document}

\title{New kids on the block: an analysis of modern blockchains}

\numberofauthors{1} 
\author{%
\alignauthor
\mbox{Luke Anderson$^\ast$, Ralph Holz$^\ast$, Alexander Ponomarev$^\dagger$, Paul Rimba$^\dagger$, Ingo Weber$^\dagger$}\\
\affaddr{$^\ast$University of Sydney, Sydney, Australia~~~$^\dagger$Data61/CSIRO, Sydney, Australia}\\
\email{$^\ast$\{l.anderson,ralph.holz\}@sydney.edu.au~~~$^\dagger$first.last@data61.csiro.au}\\
}
\date{\today}

\maketitle

\begin{abstract}

    Half a decade after Bitcoin became the first widely used cryptocurrency,
    blockchains are receiving considerable interest from industry and the
    research community.  Modern blockchains feature services such as name
    registration and smart contracts. Some employ new forms of consensus, such
    as proof-of-stake instead of proof-of-work.  However, these blockchains are
    so far relatively poorly investigated, despite the fact that they move
    considerable assets. In this paper, we explore three representative, modern
    blockchains---Ethereum, Namecoin, and Peercoin. Our focus is on the
    features that set them apart from the pure currency use case of Bitcoin. We
    investigate the blockchains' activity in terms of transactions and usage
    patterns, identifying some curiosities in the process. For Ethereum, we are
    mostly interested in the smart contract functionality it offers.  We also
    carry out a brief analysis of issues that are introduced by negligent
    design of smart contracts.  In the case of Namecoin, our focus is how the
    name registration is used and has developed over time.  For Peercoin, we
    are interested in the use of proof-of-stake, as this consensus algorithm is
    poorly understood yet used to move considerable value. Finally, we relate
    the above to the fundamental characteristics of the underlying peer-to-peer
    networks. We present a crawler for Ethereum and give statistics on the
    network size. For Peercoin and Namecoin, we identify the relatively small
    size of the networks and the weak bootstrapping process.

\end{abstract}


\section{Introduction}

Since the advent of Nakamoto's Bitcoin in 2008~\cite{nakamoto2008bitcoin}, a
diverse range of blockchain implementations have emerged. The common goal of
blockchain-based technologies is to decentralise control of a particular asset,
typically replacing one or a small set of trusted central entities with a
network of untrusted nodes. A majority of these nodes need to act faithfully to the
respective blockchain protocol in order to secure the operation of the blockchain
as a whole.

Bitcoin's asset is the Bitcoin currency (BTC) and the trusted centralised entity
it attempts to replace is the traditional bank.  Newer blockchains provide
more intricate features, such as Namecoin's attempt at providing decentralised
name registration, or Ethereum's globally distributed Ethereum Virtual Machine
that allows executing code in the form of so-called \emph{smart contracts} on
the Ethereum network. Perceptions of blockchains are varied, with opinions
ranging from highly sceptical to enthusiastic.  Success of the technology would
have broad implications for legislation, policy-making, and financial
processes, due to the inherently decentralised nature of the technology. We
believe it is too early to make a definitive call; however, blockchains are
already used to move assets on a large scale. Now is the time to investigate
them: at the time of writing, the total value of all existing bitcoin currency
equates roughly USD 7B; Ethereum's Ether follows in second place with
approximately USD 750M\footnote{Data from \url{http://coinmarketcap.com/} on
3/5/2016.}. 

In this paper, we carry out analyses for three block\-chains that have
considerably extended Bitcoin's original mechanism. Ethereum has added a
Turing-complete scripting language that makes it possible to define programs
that are executed by participants of the blockchain. Namecoin has added a
decentralised name registration service. Peercoin, finally, has made a switch
to a new consensus mechanism to ensure the security of the blockchain, namely
proof-of-stake. For the blockchains whose asset is a service that is provided
in a decentralised fashion---Namecoin and Ethereum---we focus on how the
service is used and supported by the blockchain. An analysis of Ethereum is one
of our main contributions: we show the development of the blockchain and give
fundamental statistics that describe its use. In particular, we explore the
smart contracts and uncover oddities, including some security-related (so far
low-impact) issues. We also provide a first analysis of similarity between
contracts on the blockchain, which gives hints at how they have been developed.
For the `pure-currency' blockchain, Peercoin, we investigate how much the new
proof-of-stake algorithm is involved in moving currency around. Finally, we
investigate briefly some core properties of the networks that keep these
blockchains running, including peer crawling data and tests of the
bootstrapping process.

The remainder of this paper is organised as follows. In the next section, we
provide the necessary background to understand the blockchains we investigate
and discuss previous related work. Section \ref{sec:methodology} presents our
methodology and provides an overview of our data sets. Section
\ref{sec:results} presents our results for Ethereum, Namecoin, and Peercoin. We
offer our conclusions in Section \ref{sec:conclusion}.

\section{Background and related work}
\label{sec:backgroundrelwork}

We provide the necessary background on blockchain technology in this section
and discuss related work.

\subsection{Background}
\label{sec:background}

A blockchain is an append-only, public ledger that keeps track of transactions
relating to an asset. In Bitcoin, for instance, a transaction is the transfer
of some amount of Bitcoin currency (BTC) from one account to another.  An
account is expressed as a public key; ownership of the corresponding private
key proves ownership of the account. Transactions are grouped into blocks,
which are cryptographically linked---\ie block $n$ contains a hash value of
block $n-1$.  Transactions are digitally signed: only the holder of the private
key can create a valid signature, but all participants of the blockchain can
verify its validity with the known public key\footnote{Keys are based on
elliptic-curve cryptography and are thus short and suitable for inclusion in
the blockchain.}.  All participants maintain the entire history of the
blockchain, and in most blockchains every participant knows the ownership
status of all assets and all accounts. However, one typically does not know the
identity of an account owner by only looking at this data.

A serious problem that all early crypto-currency proposals shared was the need
to maintain a trusted authority that could issue cryptographic coins and track
transactions so as to ensure that no digital coin could be spent twice, and that the
payer is able to transact if her balance permits.  One of Bitcoin's declared
goals was to remove this central authority. The key insight was that
double-spending can be prevented if all participants in the digital currency
are able to maintain the same, correct view of all transactions in the network.

\textbf{Proof-of-work} Bitcoin introduced the blockchain as a cryptographic
ledger that stores all transactions ever made. Bitcoin nodes relay transactions
to all nodes that they are connected to.  This is insufficient to prevent
double-spending, however, as an attacker could always try to partition the
network and spend the same coin independently in each partition. Hence, Bitcoin
requires all nodes to achieve consensus regarding which transactions have been made or
rejected, \eg due to insufficient balance. This is solved by
\emph{block mining}. 
Nodes collect transactions into a block and
compute a hash over the whole block, together with a nonce of their choosing. A
block will only be accepted by the entire network if the hash value,
interpreted as a number, is smaller than a given target value, which depends on
the previous blocks' hash values, and which each node can compute independently
and deterministically. 

When a miner finds a valid hash, the block header containing the valid hash is
announced to the network.  Other miners in the network then validate the new
block header before accepting it.  Acceptance is indicated by miners starting to work
on subsequent blocks, incorporating the newly mined block as the parent.  To
mine a block, a node must thus invest computational resources to find a nonce
that when used in the hashing process will yield the necessary block hash.  

It may occasionally happen that two miners find a block at the same time (a
`fork'). In that case, the blockchain is not `in consensus' until the next
block is found, which takes one of the two candidates as its chosen previous
block. The new, longest blockchain is then accepted by the network as the
`consensus view'. 

By requiring not just one, but a number of blocks to have been found since the
first inclusion of a transaction, it becomes exceedingly unlikely that an
attacker is able to forge a different, longer blockchain that contains a
different set of transactions: to change a transaction in an already confirmed
block, one must recompute the hashes for all subsequent blocks.  Bitcoin's
author suggested that an attacker would need to control strictly more than 50\%
of the network's computational power for this attack to
succeed~\cite{nakamoto2008bitcoin}, although more recent research has
identified attacks that succeed with less~\cite{Eyal2014}. 

A major assumption is that miners are incentivised to act faithfully.  As a
reward for mining a block, a miner may assign a certain amount of thus newly
generated currency to itself in the new block. In Bitcoin, the exact amount is
currently 25~BTC and this reward halves at certain intervals (52,500 blocks),
with the next halving expected to occur in July~2016.  Senders of transactions
may also choose to pay transaction fees that the block miner can claim. The
input of a transaction must be the output of a previous transaction (except for
the aforementioned currency generation). The output is one or more currency
values with the public keys of the receivers, all signed by the sender. If the
output value is lower than the input value, the miner can claim the difference;
a transaction fee.  With block rewards halving, the reward for mining will
ultimately consist of only transaction fees. 
\textbf{Mining pools} The difficulty in finding new blocks can grow rapidly
with every new block as the target difficulty is dynamically adjusted.
Participants first switched to GPUs for increased computational power (over
CPUs) and then to specialised hardware (ASICs)---but the energy costs are still
very high~\cite{odwyer2013bitcoin}. One way to participate in the network while
still realising a profit is participation in a mining pool. A mining pool is an
organisation that distributes a portion of the mining workload to individual
participants and shares the block reward amongst all participants when a block
is mined by just one participant in the pool. The drawback of a mining pool is that
it introduces a form of centralisation: particularly large mining pools may,
when they collude, hold more than 50\% of the network's computational power.

\textbf{Namecoin} Bitcoin uses a scripting language to express how a
transaction output can be claimed. Bitcoin's scripting language was extended by
Namecoin to allow name registration. A sender can invest currency (NMC) to request a
mapping of a name to some value to be stored in the blockchain. There are three
primary operations. \texttt{name\_new} announces the intention to register a
name by providing a hash value of the desired name in a transaction\footnote{A
hash value is announced to avoid another participant deciding to race for the
name.}. This transaction carries a special fee of 0.01 NMC.  A
\texttt{name\_new} is followed by a \texttt{name\_firstupdate}, which costs a
fixed fee of 0.005 NMC plus the so-called network fee, which is a quasi-linear function 
in the number of blocks and decreases over time. The network fee was
high in the beginning (50 NMC when Namecoin started on 19 April 2011) and
dropped to (rounded) zero in November 2012\footnote{The intention of the
network fee was to discourage early mass-registration.}. Names expire
(currently after 250 days) and can be renewed; they can also be transferred.
Renewal and transfer are done with the \texttt{name\_update} operation, whose fee is 0.005 NMC.
Namecoin is built off the Bitcoin code and continues to import code from the
Bitcoin repository\footnote{The last full commit that Bitcoin and Namecoin
share is from 6 May 2016.}.

\textbf{Merge mining} Namecoin also introduced merge mining whereby miners can participate in two blockchains at once. This is a
response to a persistently relatively small number of dedicated (Namecoin-only)
miners. The Namecoin software was changed to accept Bitcoin blocks as valid,
too. Miners carrying out merged mining can collect Namecoin transactions, hash
them, and include this hash value in a normal Bitcoin block. If the miner finds
a block that meets only Namecoin's target difficulty, the block is sent for
incorporation to the Namecoin network only.  If the block also meets Bitcoin's
target---which is more difficult to achieve than Namecoin's---the block is sent to both
networks, and the miner can profit twice. The Bitcoin network ignores the extra
data, but the Namecoin blockchain stores Bitcoin's data. The drawback of merge
mining is that it can introduce a strong dependency of one blockchain on
another. 

\textbf{Peercoin: proof-of-stake} Proof-of-work is wasteful as the entire
network participates in the mining process. Estimates in 2014 found the
Bitcoin network's electricity consumption to be on par with that of
Ireland~\cite{odwyer2013bitcoin}. Other consensus-building forms have been
investigated, with proof-of-stake arguably being the most interesting. The idea
of proof-of-stake is to assign the right to mine a block not based on finding a
certain hash value but on being able to demonstrate that the miner holds a
certain stake in the network, such as `coin-age', \ie a certain amount of currency
held for a certain number of days.  Peercoin uses this form of proof-of-stake.
A special function in the code determines which miners have a right to mine the
next block for a given `stake'.  A miner who mines a block resets the coin-age
for the stake to 0 and must accumulate it again before being allowed to mine a
block. Peercoin does not have transaction fees; rewards are only given out for
mining blocks. Peercoin is not entirely proof-of-stake-based: its codebase
switches over gradually from proof-of-work in Bitcoin fashion to
proof-of-stake.  The software is built off a relatively old version of
Bitcoin of June 2012\footnote{The last full commit that Bitcoin and Peercoin
share is \texttt{397737b9133118d71d2c8ba6a95afea0ba7d4350} of April 2012;
the last cherry-picked commit is from 20 August 2012.}.

\textbf{Ethereum: smart contracts} Ethereum's currency is Ether. Ethereum added
a novelty to blockchains: its scripting language is Turing-complete and allows
to define smart contracts.  These are small applications that are executed by
the entire network. The author of a smart
contract offers a certain reward for executing the code: she defines an amount
of `gas' she is willing to spend, together with an exchange rate from Ether to
gas. Smart contracts can be relatively complex; they can even instantiate other
sub-contracts and can have storage. This makes it possible to implement various
forms of contractual agreement, expressed in code. Smart contracts are written (currently) in the
high-level languages Solidity or Serpent. Mining in Ethereum currently
implements a proof-of-work model using a specialised hashing algorithm called
Ethash. Prior to the software release of Ethereum, 72 million Ether was
pre-mined and shared amongst the Ethereum foundation and early investors.
Ethereum intends to switch to proof-of-stake, which allows it to create an
unlimited supply of Ether.

\subsection{Related work}
\label{sec:relwork}

There is a body of work on Bitcoin, both for its formal properties as well as
the data stored in the blockchain.  Other blockchains seem to have received
much less attention, with Peercoin and Ethereum apparently still entirely
unexplored.

Bonneau \etal provide a description of Bitcoin's technical details and outline
major challenges and research directions in \cite{Bonneau2015}. In
\cite{Barber2012}, Barber \etal also analyse Bitcoin in depth, identify some
structural weaknesses, and propose forms of remediation.  A critical property
of Bitcoin is that it assumes miners are incentivised not to collude and
execute the protocol faithfully. In \cite{Eyal2014}, Eyal and Sirer describe
Selfish Mining, a strategy that a mining pool can follow to obtain more revenue
than its share of overall computational power suggests. The strategy does not
depend on the size of the misbehaving mining pool. The authors propose a change
to Bitcoin that does not eliminate the flaw entirely but requires a mining pool
to hold at least 25\% of the computational power of the network in order for
the strategy to succeed.  Karame \etal \cite{karame2012} were the first to
analyse the window of opportunity for double-spending Bitcoins that is open
when there is an insufficient number of confirmations to ascertain that a
transaction has been accepted by the network. Their findings highlighted how
careful one has to be in accepting Bitcoin transactions as a basis for the
transfer of real-world assets.  Decker and Wattenhofer
\cite{Decker2013} analysed the propagation of Bitcoin messages in the network
and showed the relevance of network delay to increase the chance of a
blockchain fork.  Interestingly, Bitcoin was not designed to be an anonymous
system. Although it is possible to use a different public/private key pair for
every transaction, it is still possible to match these to real-world entities
by a probabilistic argument.  Meiklejohn \etal \cite{Meiklejohn2013}
used a number of heuristics and scraping of Web sites (to obtain Bitcoin
addresses) to show that actors in Bitcoin can be identified by their
transactions and purchases they make. Ron and Shamir, finally, provide an
analysis of the Bitcoin transaction graph in~\cite{Ron2013}. Our work shares
some methodical aspects with these analyses, but focuses on the innovations
that Ethereum, Namecoin, and Peercoin introduced.

Kalodner~\etal provided a first analysis of Namecoin in~\cite{Kalodner2015}.
The authors' focus was on the use of the namespace and on the development of
an appropriate model for decentralised namespaces. They found only 28 names in
Namecoin that did not appear to be `squatted' (registered entirely for the
purpose of transferring them later for a lucrative price). Our focus in this
paper is very different: we investigate the blockchain with respect to the
merge-mining property and its development over time.

\section{Methodology and data sets}
\label{sec:methodology}

We chose the following three blockchains for our analyses: Namecoin, Peercoin,
and Ethereum. Namecoin was chosen as it was the first blockchain to introduce a
non-financial service, namely name registration, and because it was also the
first blockchain to support merge mining with Bitcoin.  Peercoin was chosen
because it was the first blockchain to introduce proof-of-stake, albeit in a
hybrid fashion. Ethereum was the first blockchain to introduce a Turing-complete
scripting language, with support for smart contracts, and is the most active
and representative blockchain of its kind.
\begin{table*}[tb]
    \centering
	\begin{tabularx}{1\textwidth}{lllrrl}
        \toprule
		Blockchain      &   First block time & Block cut-off time    & \specialcell{Block cut-off\\index}   &   No. TX      &  TX volume \\
        \midrule
        Namecoin        &   2011-04-19 12:59:40 & 2016-04-18 09:32:24   &  \num{281950}         & \num{3926035} & $1.46 \times 10^{9}$ NMC \\
		Peercoin        &   2012-08-20 04:19:16 & 2016-04-18 09:44:20   &  \num{232405}         & \num{965105}  & $1.29 \times 10^{9}$ PPC \\
		Ethereum        &   2015-07-30 15:26:28 & 2016-04-18 09:59:22   &  \num{1358548}        & \num{3470861} & $2.05 \times 10^{8}$ ETH \\
        \bottomrule
    \end{tabularx}
        \caption{Summary of blockchains studied in this paper. Note that we give the timestamp of the first \emph{conventionally} mined block. The genesis blocks, \ie block 0, were often mined days before the official release of the blockchain.}
        \label{tab:blocks}
\end{table*}
\begin{table}[bt]
    \centering
    \begin{tabular}{lrr}
        \toprule
		Blockchain      &   Size on Disk &  \specialcell{Avg. Growth \\ per Day}\\
        \midrule
		Bitcoin         &   73.0 GB & 27.0 MB\\
        Namecoin        &   3.0 GB & 1.5 MB\\
		Peercoin        &   0.5 GB & 0.5 MB\\ 
		Ethereum        &   17.0 GB & 64.5 MB\\
        \bottomrule
    \end{tabular}
        \caption{Summary of blockchain statistics.}
        \label{tab:blockstats}
\end{table}

There are two parts to our methodology. To gain a comprehensive picture of
transactions in each blockchain, we downloaded and analysed the blockchains
themselves. To gain a first understanding of the underlying peer-to-peer
networks, we analysed their bootstrapping process and carried out a crawl.

\subsection{Blockchain download}

We compiled each reference client from source, at the latest revision available
at the time our study began, and from the master branch of the repositories:
\texttt{namecoin-core}\footnote{\url{https://github.com/namecoin}, at
revision \texttt{5624a2233facea2cd6785954999b46a40b8362ac}},
\texttt{ppcoin}\footnote{\url{https://github.com/ppcoin/ppcoin}, at
revision \texttt{f01ccea4b515ce6e01f4a50cc6b50a4f3337c7ee}}, and
\texttt{ethereum}\footnote{\url{https://github.com/ethereum/go-ethereum}, at
revision \texttt{9e323d65b20c32ab1db40e8dde7172826e0e9c45}}.

For our study, we only considered blocks with a timestamp less than 18 April
09:59:59 UTC (17 April 23:59:59 \mbox{UTC-11}), but we ran each client for a
further 24 hours to ensure consensus was established. A summary of this is
provided in Table~\ref{tab:blocks}. We extracted data from the blockchains
either via the JSON-RPCs that each daemon provided (\texttt{namecoind},
\texttt{ppcoind}) or directly from the data structure on disk (in the case of
Ethereum, with \texttt{geth} as a client). Where applicable, we used the RPC
calls or provided APIs to decode or compute certain additional values (\eg
decoding Namecoin scripts or computing transaction fees). We use the PostgreSQL
database to store and query our data. Our code is available from
Github\footnote{\url{github.com/modernblockchains}}.
Table~\ref{tab:blockstats} provides an overview of storage requirements for the
blockchains we analysed. We added Bitcoin for comparison, which is by far the
largest blockchain. The two Bitcoin-based derivations Peercoin and Namecoin
need only a fraction of the space.  This is due to the much lower number of
transactions.

\subsection{Blockchain networks}
Ethereum uses a variant of the Kademlia/Kad protocol
\cite{maymounkov2002kademlia}. Instead of forwarding messages and providing a
DHT-service, the protocol is only used for peer discovery. The node distance
metrics used in Ethereum differ from the traditional Kademlia XOR metrics by
the fact that SHA3~\cite{bertoni2011keccak} is used to modify peer selection in
the discovery process.  The discovery protocol operates by transmitting a {\tt
FIND\_NODE} query with a {\tt NodeId} parameter, which is the node ID (public
key) of the client that is querying.  The receiver hashes the {\tt NodeId}
using SHA3, i.e. {\tt TargetHash} = SHA3({\tt NodeId}). Then, it computes the
hashes of all its known peers and performs the XOR operation on all the
computed hashes with {\tt TargetHash}. Finally, the receiver sorts the results
in ascending order and selects the top 16 of the peers.

We implemented a Kad crawler to gather peers using a recursive process of
issuing {\tt FindNode} queries. We modified the original \texttt{geth} client
for this to carry out the recursive queries and log all replies. This ensures
that our crawler uses valid protocol messages.

As described, a query with {\tt TargetID} will result in getting peers whose
SHA3 hash of the {\tt NodeID} starts with the same prefix as our supplied {\tt
TargetID}---if the queried peer has such peers in their database. The algorithm
for our crawler is shown in Algorithm~\ref{alg:kad_crawler}. We add a peer to
our set of known peers if we are able to exchange the so-called
\texttt{PING}-\texttt{PONG} messages---these are a necessary part of the
protocol before one can send a \texttt{FIND\_NODE} message. We log unsuccessful
attempts to establish a connection. 

Our algorithm requires the pre-computation of {\tt TargetIDs} for
a number of prefixes (we chose a 13 bit prefix to cover the address space
well).  This can be simply done by repeatedly creating an array of 64 bytes as
the {\tt TargetID} and forward-computing the value it hashes to when received
by a peer.

\begin{algorithm}[t]
	\caption{Kad Crawler Algorithm}
	\label{alg:kad_crawler}
	\begin{algorithmic}[1]
		\REQUIRE{a set of peers from local DB, \var{KnownPeers}}
		\ENSURE{\var{KnownPeers}, a set of peers}
		\STATE \var{TargetIDs} $\gets$ Precompute SHA3 as described 
		\FORALL{\var{TargetID} in \var{TargetIDs}}
		    \FORALL{\var{peer} in \var{KnownPeers}}
				\STATE send \texttt{FIND\_NODE}(\var{TargetID}) query to \var{peer}
				\STATE \var{peers} $\gets$ query response (max. 16 peers)
				\FORALL{\var{p} in \var{peers}}
          \IF{\var{p} $\notin$ \var{KnownPeers}}
  					\STATE ping-pong \var{p}
  					\IF{success}
  						\STATE \var{KnownPeers}.add(\var{p})
  					\ENDIF
          \ENDIF
				\ENDFOR
			\ENDFOR
		\ENDFOR
	\end{algorithmic}
\end{algorithm}
Namecoin and Peercoin use a much simpler peer discovery. Peers keep track of
all peers they have encountered, together with a timestamp, and connect to a
number of them. We did not investigate the networks systematically: some
information is already publicly available \cite{bitinfocharts}.  However, we
investigated the bootstrapping process for all three blockchain networks. In
the case of Namecoin and Peercoin, the official clients ship with both
hard-coded IP addresses as well as DNS names that, when looked up, return seed
IPs. To simulate a normal bootstrap process (with DNS caching effects), we
queried the DNS names several times. In the case of Ethereum, the IP addresses and
public keys for four seed nodes are hard-coded into the official client.

\section{Results}
\label{sec:results}

We first present a comparison of global statistics and then discuss our
results for each blockchain.

\subsection{Comparison of blockchains}

\begin{figure*}[tb]
       \centering
          \includegraphics[width=.8\textwidth]{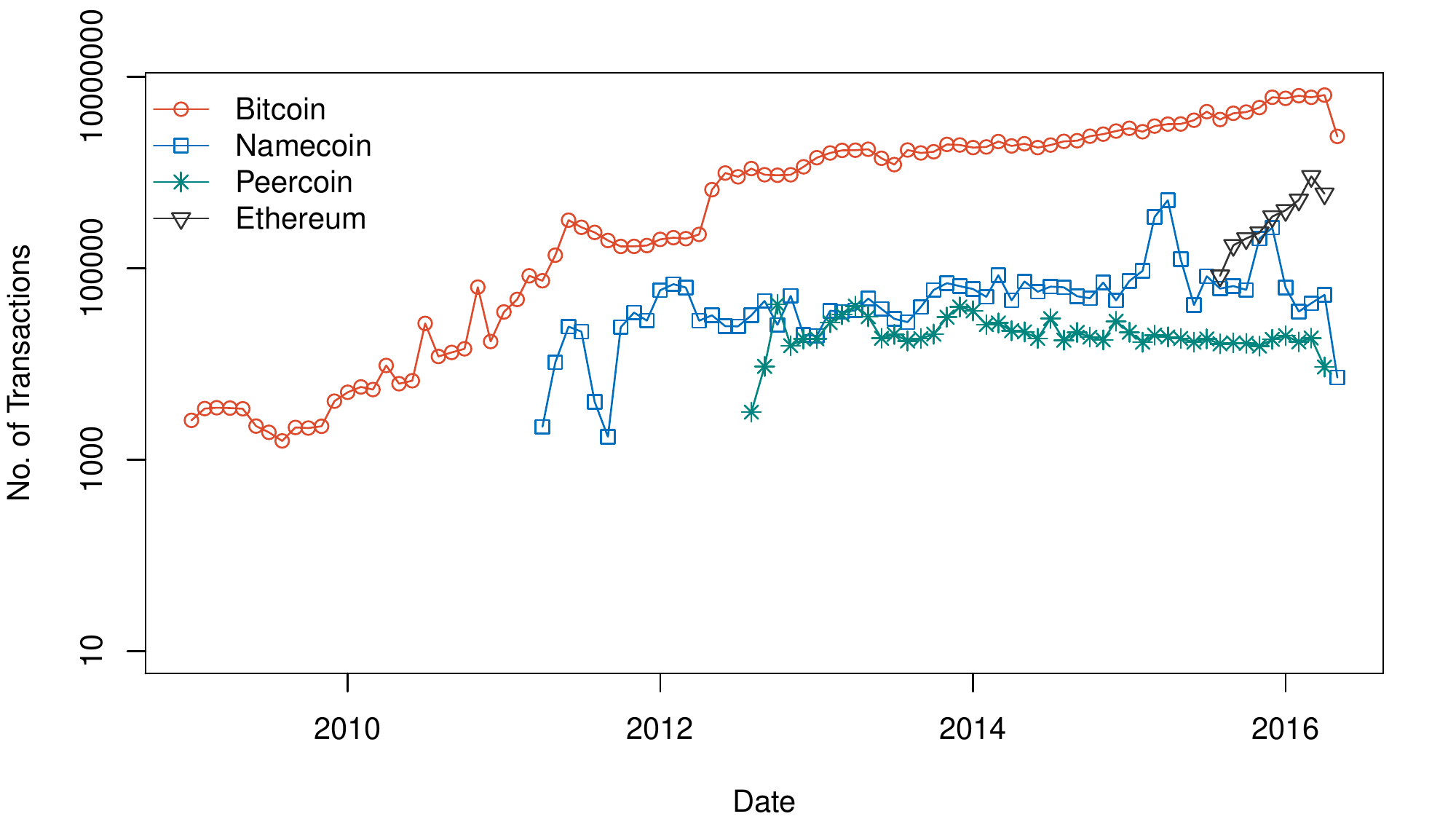}
             \caption{Number of transactions seen per month for each blockchain, plotted on a logarithmic y-axis.}
       \label{fig:blockchaintransactions}
\end{figure*}

It is instructive to analyse how our blockchains compare to each other.
Figure~\ref{fig:blockchaintransactions} shows the number of transactions that
they include in blocks (and thus accept), by month. We added data for Bitcoin
from the Coindesk public
service\footnote{\url{http://www.coindesk.com/price/}}. The figure shows that
Bitcoin is much more active (note the logarithmic $y$ axis). Starting with
meagre beginnings in 2009, it has grown to process millions of transactions a
month\footnote{The drop at the end is due to the month of April not being
completed, not a drop in transactions.}. Namecoin is almost two orders of
magnitude less active: we can identify a strong period at the beginning in
2011, followed by a drop at the end of the year and then stabilisation starting
in 2012. Despite a short peak in 2015, it has remained at roughly the same
level. Peercoin shares this fate: a strong start, and then stabilisation at a
level two entire orders of magnitude below Bitcoin. However, 
the number of transactions still translates to a serious movement of assets.
Ethereum is clearly the star among the newcomers. It started in 2015, but its
transaction count is climbing fast and it is just one order of magnitude below
Bitcoin.

\subsection{Ethereum}
\label{sec:ethereum}

We begin with an investigation of Ethereum.  Table \ref{tab:ethtransactions}
shows the number of transactions per month in Ethereum, grouped by transaction
type. Despite Ethereum being primarily intended to run smart contracts, we can
see that contract-related activity is much less pronounced than ordinary
currency transfer\footnote{Once again note that we stopped recording data on 18
April; there is no considerable drop in the number of transactions compared to
March.}. The number of transactions whose destination address is a contract
(\ie implying interaction with a contract) is approximately 15\% of the
quantity of transactions destined for regular accounts. The number of contract
creation transactions is rising, but compared to the overall transaction count
it remains small.  Evidently users are not making very good use of smart
contracts yet but use Ethereum as another currency. Note, however, that
contract creation can fail due to programming mistakes or insufficient gas in
the transaction. We discuss this below.

\begin{table}
	\begin{tabularx}{1\columnwidth}{Xrrr}
		\toprule
		Month & \specialcell{Tx. to\\Accounts} & \specialcell{Tx. to\\Contracts} & \specialcell{Tx. create\\Contracts} \\ 
		\midrule
		2015 Aug & \num{76707} & \num{5866} & \num{746} \\
		2015 Sep & \num{165117} & \num{7261} & \num{1087} \\
		2015 Oct & \num{190867} & \num{11679} & \num{1519} \\
		2015 Nov & \num{207419} & \num{26229} & \num{1265} \\
		2015 Dec & \num{212679} & \num{132081} & \num{1526} \\
		2016 Jan & \num{278566} & \num{121846} & \num{1735} \\
		2016 Feb & \num{434222} & \num{78810} & \num{3889} \\
		2016 Mar & \num{792248} & \num{113984} & \num{4551} \\
		2016 Apr & \num{522558} & \num{74675} & \num{1729} \\
		\bottomrule
	\end{tabularx}
    \caption{Transaction counts Ethereum by type and month.}
	\label{tab:ethtransactions}
\end{table}

\subsubsection{Smart contracts}
\label{SEC:contractAnalysis}

At our cut-off time, we found a total of \num{19528} contracts in the blockchain.
\num{17424} (89.2\%) were created by transactions directly, which implies
creation by an Ethereum client and hence most likely a human actor.  \num{2104}
(10.8\%) contracts were created by other contracts.  Contracts can be
terminated with a special opcode (\texttt{SUICIDE})---of the contracts we had
found, \num{18105} (92.7\%) were active, and \num{1423} (7.3\%) terminated.
The active contracts had a balance of 3.7M Ether between them. Most of the
terminated contracts had no balance left, with the exception of 260, where 98
had a minuscule balance of $\leq$ 100 Wei (1 Wei is $10^{-18}$ Ether). However,
the remaining 162 contracts have a balance of \num{5440} Ether together---with
\num{5273} Ether held by just ten. At the time of writing, 1 Ether is evaluated
at roughly USD 9, so this translates to about USD \num{45000} lost as this
Ether can never be retrieved. We return to the topic of such lost money below.

\begin{figure}[tb]
   \centering
      \includegraphics[width=0.8\columnwidth]{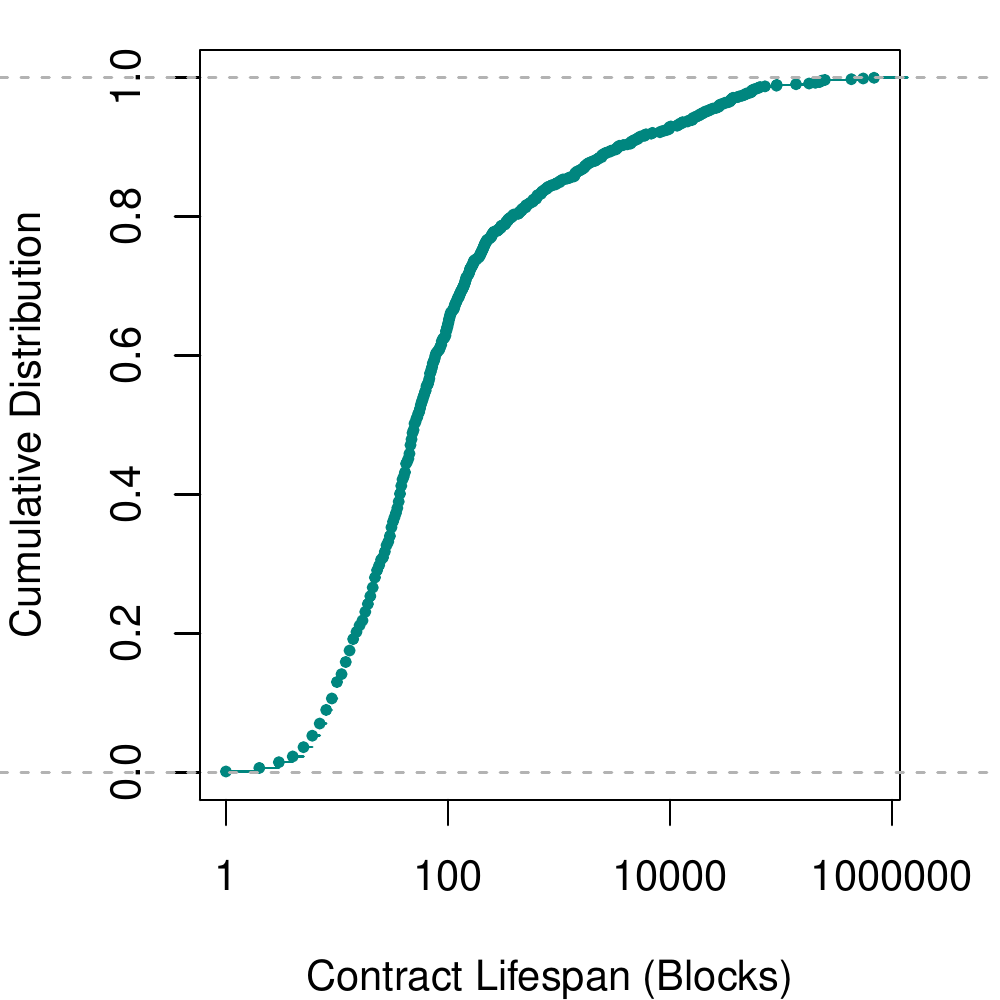}
   \caption{The lifespan of terminated contracts measured in number of blocks.}
   \label{fig:contractLifespan}
\end{figure}

We investigated the lifetimes for terminated contracts, \ie the time between
their creation and their termination. Figure \ref{fig:contractLifespan} shows
the result. The lifetime is expressed in blocks; the average time between two
blocks in Ethereum is between 14--17 seconds. As can be seen, many of the
terminated contracts had rather short lives: 60\% of them lasted for only 100
blocks, and another 20\% for about \num{10000} blocks.

\paragraph{Zombie Contracts}
\label{SEC:curiositiesZombieContracts}

A contract creation transaction is defined by setting the `to' field to
\emph{null}. Typically, this type of transaction would include some compiled
EVM code in the `input' field, which would then be executed to create a new
contract. An `endowment' of gas is often provided which provides the new
contract with a positive balance upon initialisation.  If a transaction is
submitted to the network with a blank `to' field and an empty `input' field, a
contract will be created which has no EVM code. It is safe to assume that this
is almost never intentional since these contracts lack code that defines their
operation and hence they cannot perform any actions. The Ether that the
contract was endowed with is lost forever. We call these `zombie contracts'.

\begin{figure}[tb]
   \centering
      \includegraphics[width=.8\columnwidth]{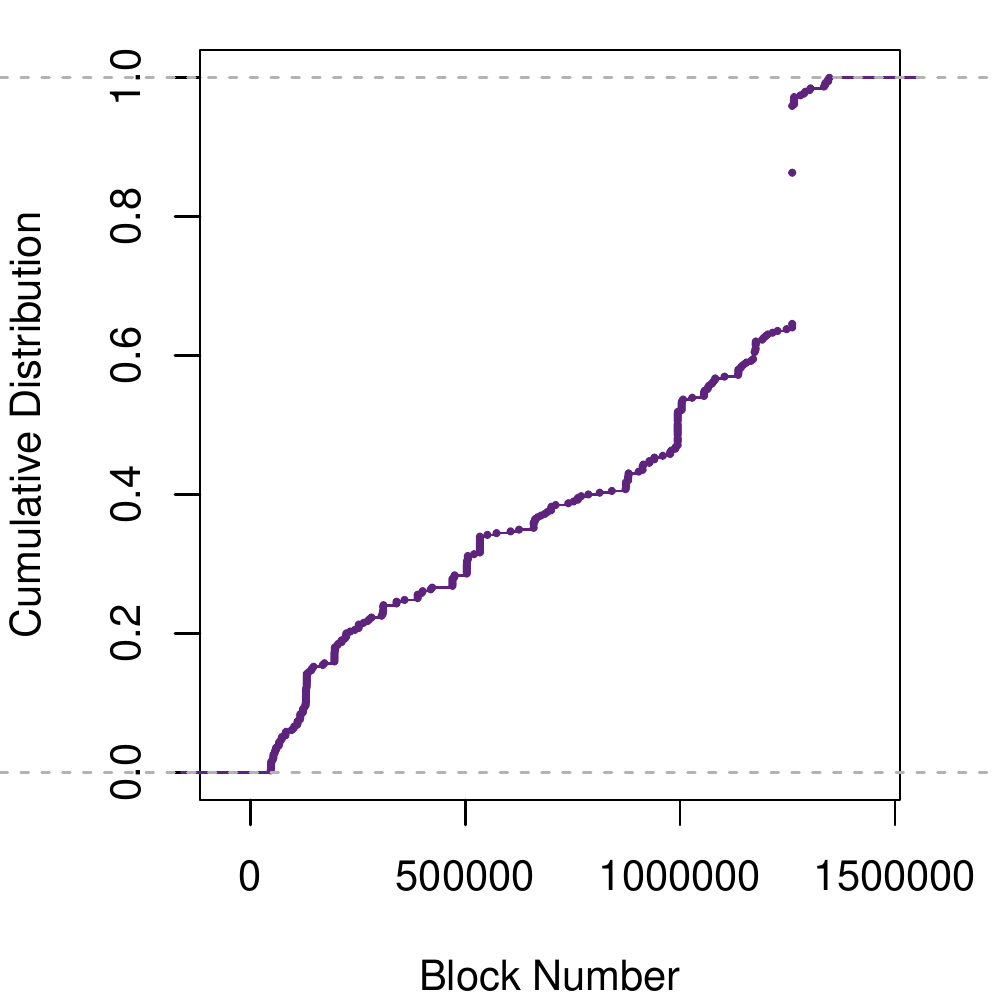}
   \caption{CDF: zombie contract creation, plotted against block number.}
   \label{fig:emptyContractCDF}
\end{figure}

We identified 395 such contracts in Ethereum. Their balances amount to
\num{5428} Ether, 97\% of which (\num{5274} Ether) belongs to the top 10
(ranked by balance). The highest balance in a single one of these contracts is
\num{2400} Ether, which at the time of writing means USD \num{21600}.  We
originally hypothesised that these errors happened more frequently in the early
days of Ethereum; however, this is not so. The median block number for
{\itshape all} contract creation transactions is \num{994493}.  Of the 395
zombie contracts, 188 were created in blocks before and including block number
\num{994493}; 207 contracts were created after. Evidently, the mistake still
happens.  Figure~\ref{fig:emptyContractCDF} shows a cumulative distribution
function of all zombie contracts, plotted by block number. The quasi-linear
distribution reinforces that this apparent mistake is a regular
occurrence throughout Ethereum's short history. 

We speculate that the zombie contracts were accidentally created by users of
{\tt geth} (a command line client) while attempting to send Ether in a
transaction. When constructing a transaction, it is necessary to enter a long
string of characters. Due to the amount of scrolling involved, this can lead to
the sender inadvertently omitting the `to' field.

An exception to the linear increase in zombie contracts is a spike around block
\num{1260882}, which is caused by 127 zombie contracts being created by a
single Ethereum address. We have no plausible hypothesis why this happened.

\paragraph{Contracts referenced before creation}
\label{SEC:curiositiesContractsReferencedBeforeCreation}

We discovered a curious pattern where a smart contract is sent Ether prior to
its creation. We found 10 examples of this pattern.  When a contract is
created, it assumes a particular Ethereum address that is subsequently used to
interact with that contract. This contract address is calculated with a
deterministic function that depends only on details of the creator's Ethereum
account (defined by equation 82 in \cite{wood2014ethereum}). It is therefore
possible to pre-compute all potential contract addresses that a particular
account is able to create.

The reason for this pattern is unclear to us. Without knowing who the address
owners are, one can only speculate. It may simply be the result of
experimentation with the blockchain; however, a more intriguing speculation is
that this is a weak form of `security by obscurity' with the intention to hide
ownership of Ether. An individual could start by generating a new Ethereum
account, then calculate the contract addresses that can be created by the new
account. Finally, she may send Ether to one of those contract addresses,
without actually creating a contract at that address.  Na\"ive inspection of
the individual's accounts would lead observers to believe they have a zero
balance. It would also allow the individual to receive Ether from other
sources, without the contract address being easily linkable to her identity.
When the individual wants to redeem the balance, she would only need to create
a contract at the address and provide code that simply pays the entire balance
to an account of her choosing.

\subsubsection{Security analysis of smart contracts}
\label{SEC:contractAnalysis-SUICIDE}
\label{SEC:contractAnalysis-SUICIDE-setup}
\label{SEC:contractAnalysis-SUICIDE-stats}

Smart contracts can be terminated with the {\tt SUICIDE} command (opcode {\tt
0xff} as specified in \cite{wood2014ethereum}), which must be called from
within a contract that is to be terminated.  A smart contract should first
check if the caller is authorised to execute the {\tt SUICIDE} command, however
responsibility for this check is left to the programmer of the smart contract.
This is an example of necessitated defensive programming in Solidity, also
described in \cite{leastauthority2015ethanalysis}.  We investigated how many
Ethereum smart contracts have an unprotected {\tt SUICIDE} command that can be
activated by an unauthorised party. For this purpose, we cloned the Ethereum
blockchain and ran a private copy that does not interfere with the real
network.

A {\tt SUICIDE} command, by the Ethereum specification, refunds all the remaining
balance that is held by a contract to a specific address. This address can be
specified by a parameter passed to the {\tt SUICIDE} command, which could be
the creator of the contract, the message sender, a hard-coded address, or a
calculated address.  A function call costs a base amount of \num{21000} gas for
the call itself, plus any additional gas that is consumed by the opcodes
present in the function. The {\tt SUICIDE} command provides a gas {\itshape
refund} of \num{24000} gas. It is one of only two commands that provide a
refund; the other is used to set a storage value in a smart contract from
non-zero to zero.  Any method that uses less than \num{21000} gas is likely to
have called the {\tt SUICIDE} command.

In order to cause contracts to execute the {\tt SUICIDE} command, it is
necessary to call a function within the smart contract that contains that
command. To call a function within a contract, one simply sends a transaction
to the contract with the `input' field set to the signature of a function. This
signature is the hash of the function name as it appeared in the source code.
Since one cannot simply read the function names from the compiled EVM code, we
first had to determine which functions could be called to achieve our desired
outcome. We did this with a dictionary attack by trying commonly used words
that may be used as function names for contract termination, e.g.  `kill',
`terminate', `destroy'.

We then analysed all active contracts and called Ethereum's {\tt estimateGas}
method with our potential {\tt SUICIDE} function calls. This allowed us to
leverage our gas consumption metric (see above) without actually sending
transactions. Any function call with a gas consumption estimate of less than
\num{21000} was considered to be a potentially vulnerable contract.  After
identifying a small subset of potentially vulnerable smart contract functions
(using our function name guessing method), we then queried historical data from
the Ethereum public blockchain to identify function signatures that were used
to terminate contracts. We then used some of these signatures to further
identify additional active contracts that are vulnerable.  After identifying a
contract that was potentially vulnerable, we invoked the function on that
particular contract and checked whether the we did indeed terminate the
contract. We also observed the account to which the contract sent its remaining
balance, if any.

We analysed the \num{18105} active contracts and found that 816 (4.5\%) are
vulnerable to at least one of the 14 function signatures that we use.

The total balance of these 816 contracts is only 0.285 Ether. No contracts
could be terminated with the method names {\tt suicide}, {\tt end}, {\tt destroy}, {\tt done},
{\tt delete}, {\tt redeem}, or {\tt terminate}. One contract had a {\tt remove} method,
but a zero balance. 754 contracts were vulnerable to the {\tt kill} method. Out
of these, 728 had a zero balance. We also identified 58 contracts that are
vulnerable to unknown function names, \ie where we only know the function
signature. 
There were three contracts that seemed vulnerable to all tested functions.
However, none of the instructions to terminate the contracts were successful.
We suspect that they implemented a default entry function. This will make them
seem vulnerable without actually being so. 

For the 813 contracts that we terminated successfully, we analysed where their
remaining balance was sent.  76 contracts refunded their balance to us, at a
total of 247 Wei. 11 contracts sent their balances to the null Ethereum
address. The remaining 726 contracts refunded their balances to their
respective creators' accounts.  Over their lifetime, these 813 contracts
participated in a total of \num{18506} transactions, with a combined
transaction volume of 7 Ether.  Only one of them was a contract-created
contract, while all others are human-created contracts.

We found that we had to pay \num{10693} gas on average to terminate a contract.
But even though the termination methods were obviously unprotected, we never
received a positive refund. In other words, there is no monetary incentive to
terminate a contract.

\subsubsection{Similarities between smart contracts}
\label{SEC:contractAnalysis-distance}

At this stage, it is plausible to assume that participants still
experiment with the technology. As such, it is reasonable to believe that they
may also test contracts from tutorials, or variants. We investigated this.

We chose nine particularly easy-to-find example contracts and compiled each of
them with and without the optimisation flag that the compiler offers. We then
calculated the Levenshtein distance to all contracts on the Ethereum
blockchain. We disregarded anything beyond a threshold of \num{1000} substitutions,
which is equivalent to 500 ASCII characters.
Table~\ref{tab:contract-levenshtein} shows the Levenshtein distance ($\delta$)
for each of the example contracts, categorising the results into exact copies,
minor changes ($0<\delta\leq 100$), and heavy modifications ($100<\delta\leq
1000$). There are 309 contracts that are exact copies of the tutorial contracts
and \num{2937} contracts that are very similar to the tutorial contracts. This
is a surprisingly large number, given the small number of contract creations.
As contracts from tutorials are generally small toy examples, the higher distances
may either indicate heavy modifications or entirely new contracts that are
similarly structured.

\begin{table*}[]
\centering
\begin{tabular}{llrrrr}
\toprule
    \multicolumn{2}{l}{\multirow{2}{*}{Tutorial Contract}}  & Size (compiled) & \multicolumn{3}{c}{Levenshtein Distance ($\delta$)}                                \\ 
	\multicolumn{2}{l}{}                                   & ~ & $\delta=0$ & $0<\delta \leq 100$  & $100 < \delta \leq \num{1000}$ \\ \midrule
    Ballot$^\ast$               & optim          & \num{2254} & 7         & 1                                & 3                                   \\
    Blind Auction$^\ast$                  & optim    & \num{2780}      & 0         & 0                                & 0                                   \\
	Safe Remote Purchase$^\ast$           & optim    & \num{1368}      & 0         & 0                                & \num{5137}                                \\
	Simple Auction$^\ast$             & optim        & \num{1106}  & 1         & 0                                & \num{6761}                                \\
\midrule
    \multirow{2}{*}{Coin$^\diamond$}      & optim          & \num{478} & 94        & 1                                & \num{8815}                                \\
                                      & non-optim      & \num{1018} & 18        & 93                               & \num{9311}                                \\ \midrule

    \multirow{2}{*}{Crowdsale$^\dagger$}  & optim          & \num{2528} & 4         & 0                                & 29                                  \\
                                      & non-optim      & \num{4094} & 1         & 0                                & 15                                  \\ 
    \multirow{2}{*}{Greeter$^\dagger$}    & optim          & \num{734} & 127       & 47                               & \num{9183}                                \\
                                      & non-optim      & \num{1086} & 39        & 11                               & \num{6776}                                \\ 
    \multirow{2}{*}{Mortal$^\dagger$}     & optim          & \num{186} & 7         & 48                               & \num{8669}                                \\
                                      & non-optim      & \num{432} & 11        & 7                                & \num{8827}                                \\ 
                                      
    \multirow{2}{*}{Token$^\dagger$}      & optim          & \num{146} & 0         & \num{2728}                             & \num{5965}                                \\
                                      & non-optim     & \num{390} & 0         & 1                                & \num{8906}                                \\ \midrule
\end{tabular}
    \caption{Levenshtein distances of contracts to publicly available tutorials (compiled form). Contracts marked with $^\ast$ are from \url{solidity.readthedocs.io}, $^\diamond$ from \url{ethereum.gitbooks.io}, and $^\dagger$ from \url{ethereum.org}.}
    \label{tab:contract-levenshtein}
\end{table*}

\subsubsection{Network}

\begin{figure*}[tb]
   \centering
      \includegraphics[width=.75\textwidth]{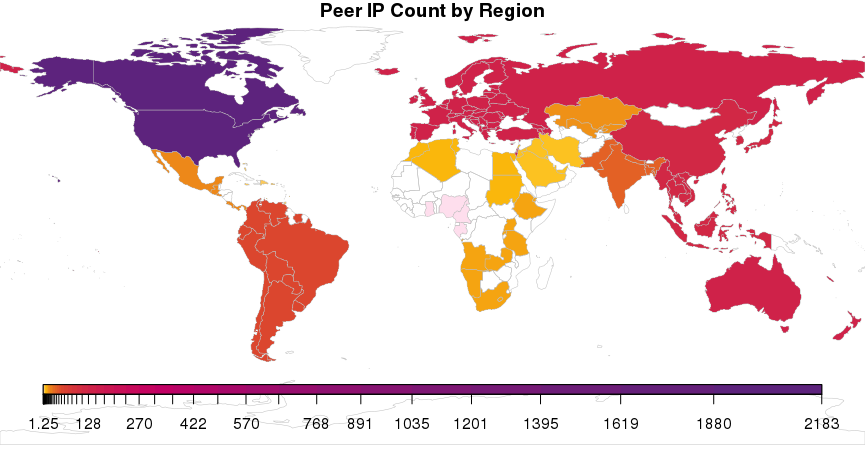}
	  \caption{Distribution of Ethereum node IP addresses by country}
   \label{fig:ethereumnodedist}
\end{figure*}

As described in Section~\ref{sec:background}, Ethereum uses a peer discovery
protocol that is modelled on Kademlia.
Ethereum nodes
are specified by public keys, called `node IDs', which are used to secure
communications. 

\paragraph{Bootstrapping}

The official Ethereum client, {\tt geth}, comes with four hard-coded IP
addresses that serve as bootstrap nodes for the P2P network.  Their public keys
are also hard-coded to prevent Eclipse attacks \cite{heilman2015eclipse} upon
initial bootstrapping, where the attacker controls the paths to IP addresses
that are vital for the peer's integration into the network and tries to
impersonate the bootstrap nodes.  We scanned the four IP addresses to determine
their availability. The first three {\tt geth} clients were available for
connection on the Ethereum standard port (30303) via both TCP and UDP. They
appeared to be running on AWS infrastructure based in Ireland, S\~ao Paulo, and
Singapore. The fourth client appears to be located at a German IP address and
was not available for connection on the standard Ethereum port.

Since 3 of 4 seed nodes appear to be operating on stable infrastructure (AWS),
they are not likely to become unreachable. However, denying a client access to
these three IP addresses would prevent automatic P2P bootstrapping.
Bootstrapping via DNS would be an addition; however, depending on his position
in the network, the attacker could try to modify the DNS replies.

\paragraph{Crawling}

We ran our crawler, described in Section~\ref{sec:methodology}, for three days,
starting on 2016-05-06 and ending on 2016-05-09. We crawled from our host in
Sydney, Australia, which features a 100Mbit/s uplink to the Australian Research
Network. The hardware was a quad-core Xeon at 2.4GHz, with 64 GB of RAM. We
capped it at a maximum of 500 concurrent \texttt{FIND\_NODE} queries. Table
\ref{tab:ethcrawl} summarises our findings.

\begin{figure}[tb]
   \centering
      \includegraphics[width=\columnwidth]{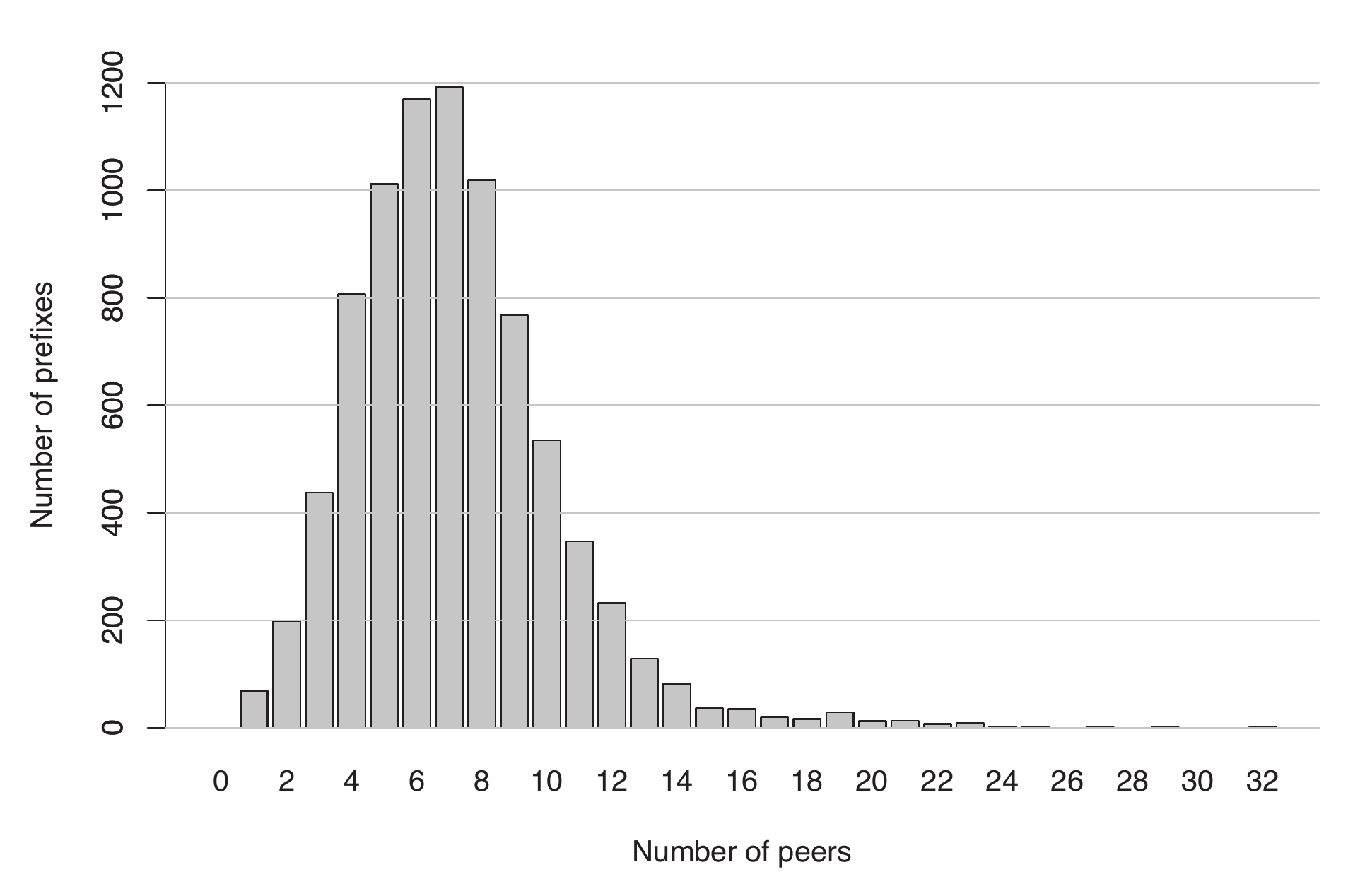}
    \caption{Number of prefixes ($y$-axis) containing a certain number of peers ($x$-axis).}
   \label{fig:ethereum-crawler-peerdist}
\end{figure}

Over the three days, we collected \num{58612} unique NodeIds from the responses
to our {\tt FIND\_NODE} queries.  As Ethereum clients generally do not change
their NodeIds (unless reinstalled), this gives us an estimate of the number of
peers in the network. Figure~\ref{fig:ethereum-crawler-peerdist} shows how many
13-bit prefixes ($y$-axis) contained a certain number of peers ($x$-axis).
Although it may seem nearly normal, there is a long tail. One would expect a
uniform distribution of PeerIds over the address space; this does not seem to
be the case. Table \ref{tab:ethereum-crawler-peerRegistered} shows a further
peculiarity: just five IP addresses were returned for over \num{40000} PeerIds,
with one IP from a German hosting provider being responsible for almost
\num{36000} PeerIds alone. It is hard to determine what the reason for this
might be; but running such a high number of PeerIds on one machine alone could
be indicative of an attack on the network. However, as Ethereum's Kad variant
is not used for forwarding messages and there are more nodes available for
discovery, this seems not too plausible here.

Mapping the NodesIds to IP addresses and ports,  we found a total of
\num{146091} combinations of IP and port, but only \num{26378} distinct IPs.
During the crawl, we received valid replies from only \num{9215} unique IP
addresses. Many clients are not reachable from the public Internet. In fact,
\num{1196} IP addresses in our result were from the ranges 172.16.0.0/12,
192.168.0.0/16, and 10.0.0.0/8---although only 50 of them were distinct.  A total of
\num{51792} distinct ports were in use---once again a sign of peers being
behind NATs. The standard port, UDP/30303, was found \num{26317} times in the
IP:port combinations. This has only a small impact on the blockchain, however:
peers that are behind NATs cannot be freely contacted by others to receive
relayed transactions.  They rely entirely on the number of connections they
make themselves. However, they are free to make many such connections. Ideally,
they should choose peers with low latencies; this does not seem to be
implemented yet, however.

\begin{table}[tb]
    \centering
    \begin{tabular}{lr}
        \toprule
        IP:Port combination &   \num{146091} \\
        Unique IPs          &   \num{26378} \\
        Unique Ports        &   \num{51792} \\
        IPs in private ranges & \num{1196} \\
        \bottomrule
    \end{tabular}
    \caption{Key statistics from our Ethereum network crawl.}
    \label{tab:ethcrawl}
\end{table}

\begin{table}[th]
\centering
\begin{tabular}{@{}lll@{}}
\toprule
\textbf{IP address} & \textbf{\begin{tabular}[c]{@{}l@{}}Unique NodeIds \\ registered\end{tabular}} \\
\midrule
78.46.49.102        & \num{35943}                                                \\
24.65.141.220       & \num{1565}                                                 \\
62.176.104.144      & 743                                                  \\
115.66.178.58       & 650                                                  \\
188.166.179.233     & 29                                                   \\
\bottomrule
\end{tabular}
\caption{IP address with the highest number of unique NodeIds in Ethereum.}
\label{tab:ethereum-crawler-peerRegistered}
\end{table}

We were also interested in understanding the geographic distribution of peers
in the Ethereum blockchain. We resolved the IP addresses we had obtained
with the Maxmind GeoLite database of May 2016. Figure
\ref{fig:ethereumnodedist} plots the distribution on a world map. The dominance of
Western countries, China, and Russia is evident.

\subsection{Namecoin}
\label{sec:namecoin}

Namecoin is different from Bitcoin in two important aspects. First, it allows
merge-mining.  Second, it allows the
registration of names in the blockchain.

\begin{figure*}[tb]
   \centering
      \includegraphics[width=.75\textwidth]{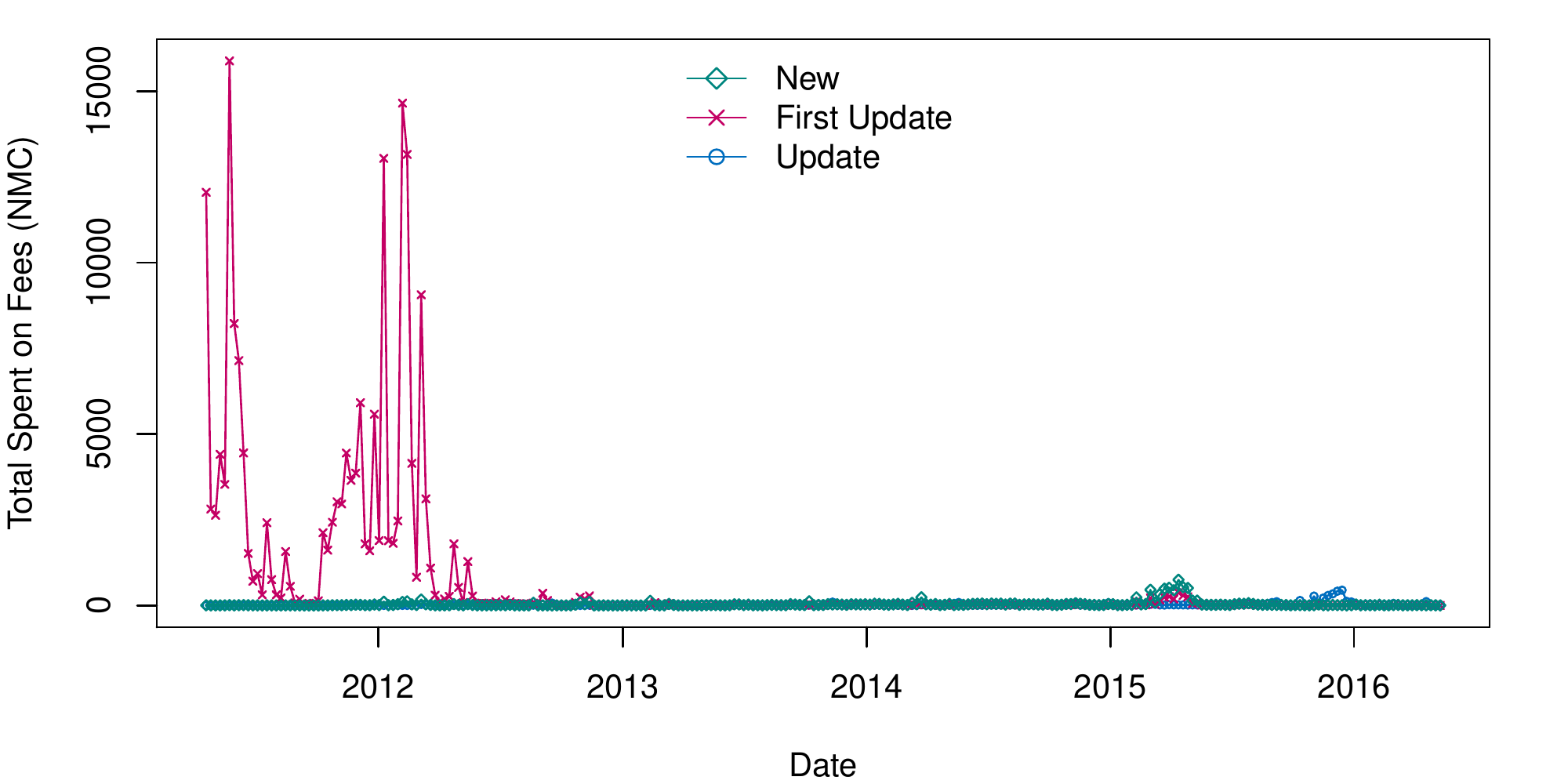}
      \includegraphics[width=.75\textwidth]{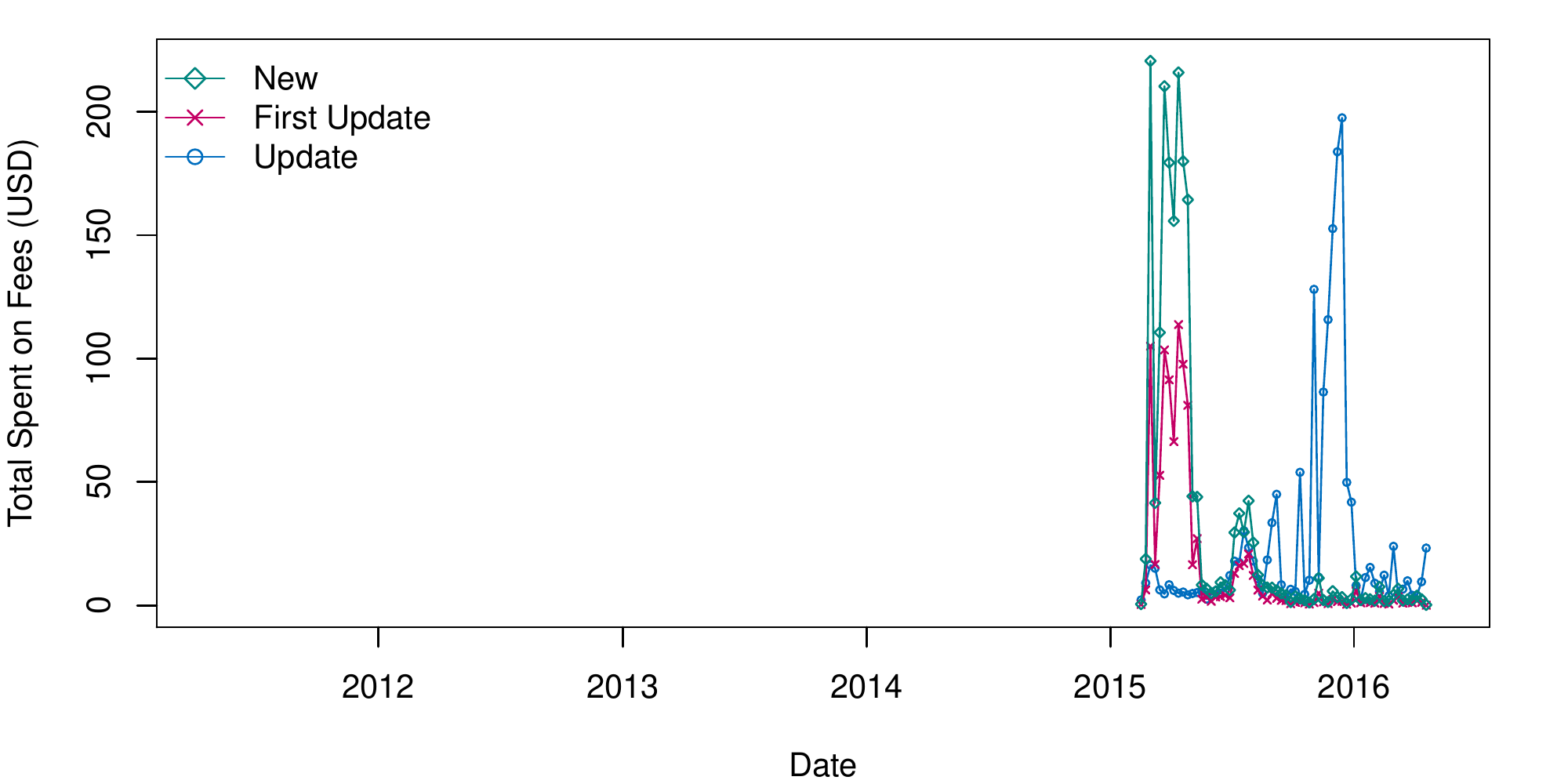}
	  \caption{The weekly sum of transaction fees collected for \texttt{new},
	  \texttt{first\_update}, and \texttt{name\_update} transactions. The top and bottom charts show the fees in Namecoins and US Dollars respectively (only partial USD data is available due to limitations of the source: \url{https://www.poloniex.com/}{Poloniex}).}
   \label{fig:namecointransactionfees}
\end{figure*}

The total number of Namecoin blocks included in our study until our cut-off
time was \num{285077}. Of these, 93\% were merge-mined (\num{265747}). Given
that merge-mining was only introduced with block \num{19200}, this means that
since that block---mined on 8 October 2011---only 130 true Namecoin blocks were
mined. All other blocks were mined by Bitcoin miners that support merge-mining
with Namecoin. This demonstrates an enormous dependency of Namecoin on
Bitcoin, which is also reflected in the number of transactions: just 1.7\%
(\num{67513}) of them are in classic Namecoin blocks while the remaining 98.3\%
(\num{3.9}M) are in merge-mined blocks. Concerning name operations, we find
that more than 99\% of these are in merge-mined blocks.
Table~\ref{tab:namecoinauxpow} summarises this. It is fair to say that the
Namecoin blockchain presently receives almost its entire computational power
from merge-mining with Bitcoin.

The implications of this can be discussed controversely. On the one hand,
adding Bitcoin's power to Namecoin means that any attack on the name
registration system requires an inordinate amount of computational resources.
On the other hand, it means that Namecoin is essentially a kind of `side-chain'
of Bitcoin. Since merge-mining implies that entire Bitcoin blocks must be
written into the Namecoin blockchain, there is also an extreme overhead of
storage required in Namecoin.  Recall that the Namecoin blockchain requires 3GB
on disk, and compare this to the requirements of Peercoin.  Peercoin has a
third of the transactions of Namecoin, but requires only a sixth of the space
that the Namecoin blockchain needs. The storage overhead for Namecoin is close
to 100\%.

\begin{table*}
    \begin{tabularx}{1\textwidth}{lccccc}
        \toprule
        ~                               & ~                &   ~    &   \multicolumn{3}{c}{Name operations} \\
        ~                               & Blocks           &   Transactions               &\texttt{new}        &   \texttt{firstupdate}        &   \texttt{update} \\
        \midrule
        Normally mined                  & \num{19330}           &  \num{67513}      &   \num{5484}          &   \num{2817}                  & \num{2624} \\
        Merge-mined                     &\num{265747} (93.2\%)  &\num{3900753} (98.3\%) & \num{964778} (99.4\%) & \num{867733} (99.7\%) & \num{1081790} (99.8\%)\\
        \bottomrule
    \end{tabularx}
    \caption{Activity in Namecoin, separated by normally mined blocks and merge-mined ones.}
    \label{tab:namecoinauxpow}
\end{table*}

It is instructive to inspect if Namecoin is used in some sensible way.  An
initial analysis of Namecoin was provided by Kalodner \etal
in~\cite{Kalodner2015}. The authors report that they could only identify 28
domains that did not seem to be registered by squatters for the purpose of
transferring or selling them later. We refer the reader to their publication
for details as we did not attempt to replicate their study. In the following,
we are more concerned with how the usage of Namecoin has developed over \emph{time}. At the
time of writing, we find \num{644592} registered names, and there are
\num{6066222} unique addresses (public keys) in Namecoin, distributed over
approximately 4M transactions.

Recall that registration of names costs fees in Namecoin. Figure
\ref{fig:namecointransactionfees} shows these fees over time (plotted per
week), for each type of name operation. The high network fees in the beginning
for the \texttt{name\_firstupdate} are the cause for spikes for this operation.
The costs for registrations became very small---near 1 NMC---in recent years.
Two bumps in the plot indicate sudden surges of registrations---but apart from
this, there does not seem to be too much activity. This is also reinforced by
the second plot in Figure \ref{fig:namecointransactionfees}. We used the
exchange rates provided by the Poloniex currency
exchange\footnote{https://poloniex.com} in its API to convert Namecoin values
to Bitcoin and from there to USD. One can see that investment per week topped
out at 200 USD during spikes, but has generally been very low.

We analysed two of the spikes that were prominent in April 2015, namely on
2015-04-11 and 2015-04-25. Our hypothesis was that these might be renewals of
expiring names.  We first checked for \texttt{name\_update} operations. We
found that only \num{1282} of the name operations on 2015-04-11 belonged to
that category, and only \num{365} on 2015-04-25. When checking the
\texttt{first\_update}s for 2015-04-11, \ie registration of a name that is not
in the system (any more), we found \num{10890} names registered on that day
(for 2015-04-25, we found \num{9049}). Of these, 170 had been previously
registered, mostly in 2014 and before. These were thus re-registrations of
expired names. The names themselves consisted of nouns, adjectives, and a
person's first name. With the exception of just four names, all were
re-registered again after 2015-04-11.  We repeated the exercise for 2015-04-25
and found very similar numbers, both absolute and relative (although the names
were more entertaining, \eg \texttt{thepiratesbay}, \texttt{starwar}
(\emph{sic!}), and \texttt{disney-world}. The conclusion that seems most likely
here is that this is a form of the squatting behaviour that the authors of
\cite{Kalodner2015} had also found.

\subsubsection{Network}

We investigated how the Namecoin blockchain is supported by its own network.
We note that the website BitinfoCharts \cite{bitinfocharts} shows roughly 200
Namecoin `active' nodes, although it is not clear from the site how `active' is
defined. 

The Namecoin software is shipped with 16 hard-coded IP addresses, plus two
Onion addresses for use over Tor. Five of them did not have reverse DNS
entries.  Of the remaining 11 names, we could identify two to be from networks
for private customers. We could identify five as hosted/co-located. One IP
belonged to a mining pool.  The rest were second-level domains in COM/NET/ORG.
We did not check activity for Onion addresses.

Namecoin also hard codes five DNS names whose A records can be queried for
bootstrap nodes. Of these, only four replied with DNS records.  We queried each
domain ten times. On every occasion, two replied with 26 A records each, one
with 32, and one with 17. The overlap was relatively small, at about 10\% of IP
addresses that were returned by more than one source. A round-robin or random
selection was also evidently taking place as we collected new IP addresses with
every run: after two runs, we had 137 unique addresses, after five runs 207,
and after ten runs 337. We queried their PTR records and identified about 180
IP addresses for private customers, 4 AWS IPs, and 26 addresses from colocation
providers.

In total, we had obtained 347 seed IP addresses from both the hard-coded and
DNS seeds. Scanning these addresses, we found 126 (36\%) of them had the
Namecoin port (8334) open, but in 183 cases (53\%) the IP seemed to be behind a
firewall, and in 39 cases the port was closed (11\%).

Our conclusion is that the Namecoin network profits from DNS-based
seeding, where servers evidently keep track of a relatively large number of
other Namecoin clients that they reveal on subsequent DNS queries. A
significant number of revealed IP addresses seem to be from private households,
however, and only about a third of them have an actual Namecoin client running
on that port.

\begin{figure}[t]
   \centering
      \includegraphics[width=0.8\columnwidth]{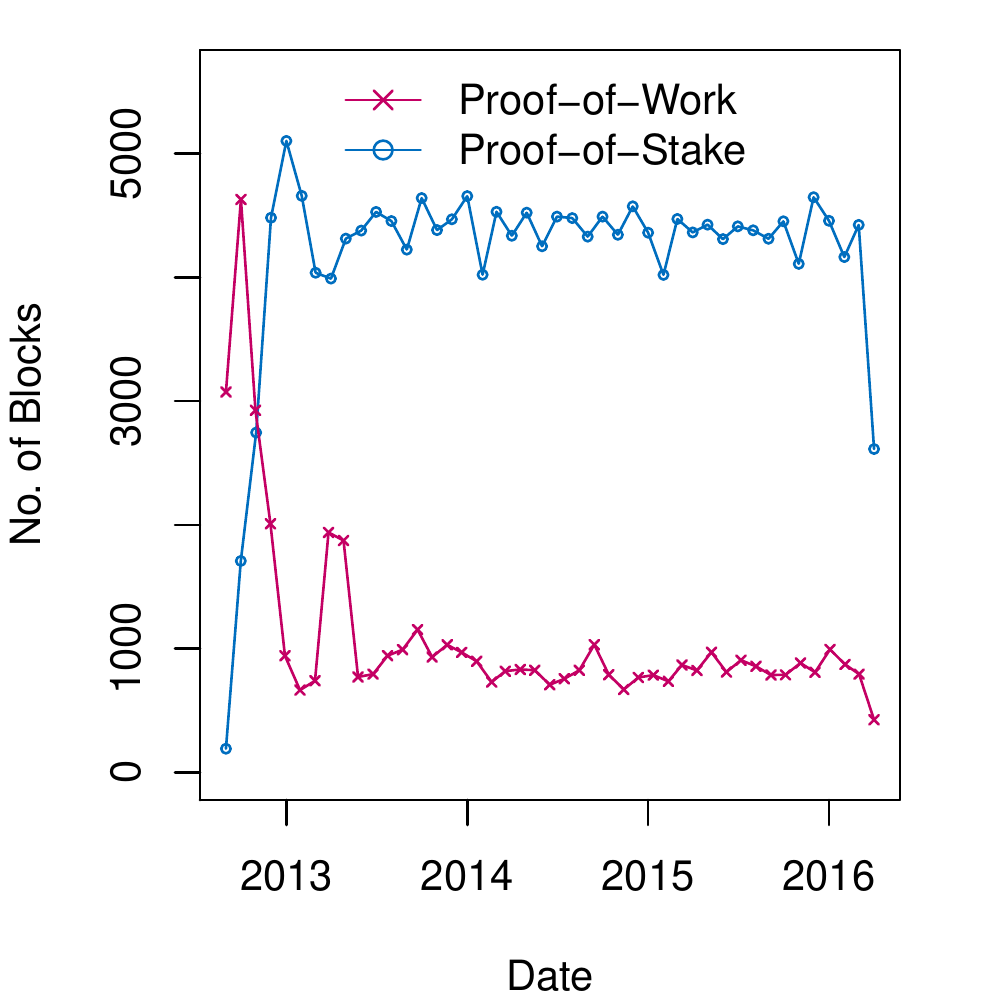}
   \caption{A comparison between the number of Proof-of-Stake and Proof-of-Work blocks in Peercoin.}
   \label{fig:ppc-pos-vs-pow}
\end{figure}
\subsection{Peercoin}
\label{sec:peercoin}

We also briefly investigated the Peercoin blockchain. Here, our primary
interest was to determine the exact usage of proof-of-stake. This concept is
described in rather vague terms in the project's whitepaper \cite{King2012}.
The whitepaper---which is from 2012---also announces that the network will
gradually move to proof-of-stake. Proof-of-work blocks will continue to play a
role, but their importance decreases.  Figure \ref{fig:ppc-pos-vs-pow} shows
how many blocks were mined by proof-of-work and how many were created with
proof-of-stake. The switch-over has, in fact, happened at an early stage of the
network, and since ca. 2013 proof-of-stake blocks constitute the vast majority.

We were also interested to see how large the network was.  There are currently
\num{638495}, which is an order of magnitude less than in Namecoin. This
corresponds well with the lower number of transactions that we found for this
blockchain. The website BitinfoCharts \cite{bitinfocharts} gives an estimate of
just under \num{1000} nodes.

For bootstrapping, 15 IP addresses are hard-coded seeds in the Peercoin client.
Peercoin also ships with 7 hard-coded domain names. We queried these for their
A records, six times in total. Each time, we received 2 NXDOMAIN. Four times, a
domain yielded a SERVFAIL, although it did respond in two cases. This was one
of the domains that we identified to return IP addresses in a round-robin
fashion. After two tries, we had obtained 71 unique IP addresses. This rose to
92 after four tries and 100 after six tries.  The vast majority (around 30
addresses each try) were provided by just two seed domains.  Together with the
hard-coded addresses, we now had 113 seed IPs.  For 95 of them, we could obtain
the reverse DNS entries. 69\% of them belonged to customers of ISPs; of the
rest, almost half could be clearly identified as co-located servers. The
remainder were second-level domains where it was unclear how they were hosted.
We found one IP address that belonged to the cryptocurrency exchange Nixmoney.
We scanned all 113 seed IPs to test whether they responded on the Peercoin port
(TCP/9901). 82 had the port open, in 26 cases it was filtered (indicating the
presence of a firewall), and in the remaining cases closed. None of the 15
hard-coded IPs responded to our probes on Peercoin port (12 filtered, 3
closed).

Compared to Namecoin, Peercoin bootstrapping depends evidently more on
transient setups by private users.  For all practical purposes,
bootstrapping into Peercoin is limited to just two domains, making Peercoin the
network with the poorest bootstrapping mechanism.

\section{Summary}
\label{sec:conclusion}

In this paper, we have studied the characteristics of modern blockchains.  We
chose three representatives of these: Ethereum as the first and very successful
blockchain that allows to run Turing-complete programs on the blockchain,
Namecoin as an attempt to base a name registration service on Bitcoin
technology, and Peercoin as the first blockchain to make use of a
proof-of-stake mechanism.  We found that Ethereum is by far the most active
blockchain gaining in popularity. However, most transactions are still
transfers from one account to another, and smart contracts are only slowly
created. A good part of them is based on widely available tutorials.
Furthermore, there are some security risks involved when defensive programming
guidelines are not adhered to---although there does not seem to be a strong
financial motivation for an attacker yet.  We also found curious transaction
patterns that may be used to hide the balance of accounts. Ethereum nodes are
distributed over the globe, with particularly many nodes in Western countries,
but also in Russia and China.  In the case of Namecoin, we found that the
blockchain shows very little activity.  Previously reported name-squatting
seems to be going on, but even this is at a low rate, despite very cheap
registration costs. Peercoin, meanwhile, has switched to proof-of-stake long
ago. Despite there being no formal analysis of its proof-of-stake mechanism, it
does move considerable assets around, albeit two orders of magnitude less than
Bitcoin. Both Namecoin and Peercoin have very few clients; we showed
furthermore that their bootstrapping process relies on unavailable seed nodes.


\bibliographystyle{plain}
\bibliography{blockchain_paper}

\end{document}